\documentclass[11pt]{article}

\usepackage{amsfonts}
\usepackage{epsf}
\usepackage{graphicx}
\usepackage{psfrag}
\usepackage{color}

\newcommand{\rme}{\mathrm{e}}
\newcommand{\rmi}{\mathrm{i}}
\newcommand{\rmd}{\mathrm{d}}
\renewcommand{\qquad}{\hspace*{25pt}}

\begin{document}

\begin{center}
{\Large{\bf New Deformed Heisenberg Algebra\\ from the $\mu$-Deformed Model of Dark Matter}}

\vspace*{0.5cm}
{A.M. Gavrilik${}^{1*}$, I.I. Kachurik${}^2$, A.V. Nazarenko${}^1$}

\vspace*{0.3mm}
{\small
${}^1$Bogolyubov Institute for Theoretical Physics, Kyiv 03143, Ukraine \\
${}^{2}$Khmelnytskyi National University, Khmelnytskyi 29016, Ukraine\\
${}^{*}$e-mail: {\it omgavr@bitp.kiev.ua}}
\end{center}

\begin{abstract}
Recently, the $\mu$-deformation-based approach to modeling dark
matter, which exploits $\mu$-deformed thermodynamics, was extended
to the study of galaxy halo density profile and of the rotation
curves of a number of (dwarf or low brightness) galaxies. For that
goal, $\mu$-deformed analogs of the Lane--Emden equation (LEE) have
been proposed, and their solutions describing density profiles
obtained. There are two seemingly different versions of
$\mu$-deformed LEE which possess the same solution, and so we deal
with their equivalence. From the latter property we derive new,
rather unusual, $\mu$-deformed Heisenberg algebra (HA) for the
position and momentum operators, and present the $\mu$-HA in few
possible forms (each one at $\mu\to0$ recovers usual HA). The
generalized uncertainty relation linked with the new $\mu$-HA is
studied, along with its interesting implications including the
appearance of the quadruple of both maximal and minimal lengths and
momenta.

\end{abstract}

\section{Introduction}


The suggestion of the existence of minimal nonzero (uncertainty of)
length linked with generalized uncertainty principle (GUP) or
relation (GUR) has been advanced in the context of string theory and
quantum gravity~\cite{GM88,ACV,Scar,AS,Maz}, see also~\cite{CLMT} and
the reviews~\cite{Garay,Hoss}.
  It was shown to 
  follow from modified or deformed
  extension~\cite{KMM} of the Heisenberg algebra (HA).
  It is worth to mention that the concept of maximum observable  momenta can play
  as well important role, see e.g.~\cite{ADV}. Such a quantity 
  was predicted, in particular, within the doubly special relativity
  theory suggesting rather simple (with terms linear and quadratic in
  momentum) modification of the right hand side of
  commutators~\cite{MaSm1,MaSm2}.
  Further it became clear that besides such a minimal extension of
  the original HA, a lot of generalizations are possible, suggesting diverse
  ways to generalize (or deform) the HA.
  As a tools to classify diverse forms of deformed HA, the concept
  of deformation function(s) is of importance, see e.g.~\cite{Saa,Jann,GKR,DN,MNT,GK2}.
  Clearly, the choice of such function must determine the corresponding GUR.
  As usual, most of the authors deal with position-momentum commutation
  relations of deformed HA that involve particular function of $X$, $P$ and
  deformation parameter(s) in its right hand side~\cite{Jann,DN,MNT}.
  It is also possible that both the right and left hand sides of defining
  commutation relation are appropriately deformed~\cite{GK}, although
  such approach may be overlapping with the case of standard commutator
  and the terms containing $X P$ and $PX$ in its right-hand side as it was
  considered in~\cite{QT}.

  In the present paper, a special form of deformed HA will be
derived in the context of the so-called $\mu$-deformation based
approach aimed to model~\cite{GKKN,GKK} basic properties of dark
matter that surrounds dwarf galaxies, and its consequences analyzed.

 The case of GUR related with the minimal $(\Delta X)_{\rm min}$ is
 the best known and well studied one.
 In relation with this, due to the conjugated roles of position and momentum,
 the concept of $(\Delta P)_{\rm min}$ has appeared.
 As it was demonstrated in~\cite{Kempf2}, a single theory --- single extended or
 generalized HA (GHA) and the corresponding GUR do exist which can jointly
 accommodate the both special quantities, $(\Delta X)_{\rm min}$ and $(\Delta P)_{\rm min}$.

Then, an interesting question arises whether it is possible that
the opposite concept of {\it maximal} uncertainties for the momentum
and/or the position does exist. Quite recently, it was shown in some
papers that such a possibility indeed can be realized~\cite{Per,Pedram,SP,Prama,HL,BB2,BB}.
Moreover, as a generalization of the already mentioned unified treatment of A.~Kempf, in the work~\cite{Per}
of  L.~Perivolaropoulos, it was explicitly shown that one can provide a theory (based on appropriate
generalization of HA) which  incorporates the whole quadruple of
 $(\Delta X)_{\rm min}$,\ $(\Delta P)_{\rm min}$,\ $(\Delta P)_{\rm max}$,\ and $(\Delta X)_{\rm max}$.

 Usual treatments in the most 
 of papers are in a sense model-independent, implying a kind of universality.
 That means, physical meaning of $(\Delta X)_{\rm min}$,\ $(\Delta P)_{\rm min}$,\
 $(\Delta P)_{\rm max}$,\ and $(\Delta X)_{\rm max}$  is rather universal
 and depends on Planck length or its inverse i.e. Planck energy scale (Planck mass).
On the contrary, our treatment is based on (related with) special
deformed HA deduced in the framework of particular model of dark
matter. It is remarkable that all the four quantities:
 $(\Delta X)_{\rm min}$,\ $(\Delta P)_{\rm min}$,\
 $(\Delta P)_{\rm max}$,\ and $(\Delta X)_{\rm max}$ do appear.
So it is clear and natural that the physical meaning of this
quadruple is tightly linked with physics of the model, i.e. with
properties of the halo of DM hosted by dwarf galaxies.

For our case (connection with DM) some motivation was due to the
work~\cite{Per}, since therein the cosmology-related uncertainty
relation was explored, along with  
clear meaning of maximal length:  as suggested in~\cite{Per}, 
 this quantity can be naturally interpreted as {\it cosmological horizon}.

The uncertainty relation in its initial form due to Heisenberg is
linked with the standard commutation relation and {\it is shared} by
different states. Unlike, for all the deformed versions of HA,
explicit dependence of GUR on particular state does appear -- for
deformed oscillators this was noticed in the pioneer
papers~\cite{Bied,Macf}. 
 In our present paper, just this fact/property is in the focus and
 exploited to full extent.

Unlike the approach perceived in~\cite{Harko} and some other papers
also exploring galaxy rotation curves with the use of the well-known
Lane--Emden equation (LEE), in our line of research we deal with the
($\mu$-)Bose-condensate model of dark matter~\cite{GKKN}, and with
such tool as $\mu$-deformed analogs~\cite{GKK} of LEE. In general,
as it is well-known, deformation of an object under study is not
unique, and in~\cite{GKK} we encountered two different possible
forms of $\mu$-deformed LEE, with the corresponding different sets
of solutions, one of which being the deformed function $\sin_\mu(kr)/(kr)$.
In the present work, the third form of LEE will be
introduced that nevertheless possesses the indicated solution as
well.  Just from the requirement of equivalence of two seemingly
different deformed versions of LEE, the new $\mu$-deformed HA can be
deduced and its basic properties and consequences explored.

The paper is structured as follows. In \textbf{Section~\ref{Sec2}},
some basics of $\mu$-deformation and $\mu$-deformed calculus are
presented. In \textbf{Section~\ref{Sec3.1}} we describe relevant
deformed analogs of LEE and, from the condition of their
equivalence, obtain the $\mu$-analog of HA which is the central
object of this work. The corresponding GUR which involves the
parameter $\mu$ is derived, and its main properties are explored in
\textbf{Section~\ref{Sec3.2}}, including the appearance of minimal
and maximal uncertainties of both position and momentum.
\textbf{Section~\ref{Sec3.3}} is devoted to discussion of
implications of these quantities for dark matter. The paper is ended
with concluding remarks.

\section{Deformed Functions and Calculus}\label{Sec2}

\subsection{Basis Functions}

The so-called $\mu$-bracket of a number or operator $X$,
\begin{equation}
[X]_\mu\equiv\frac{X}{1+\mu X}; \qquad \ \ [X]_\mu \to X, \
{\rm if} \ \mu\to 0,
\end{equation}
and the related $\mu$-deformed oscillator have been introduced three
decades ago in~\cite{Jann}. More recently, there appeared some
papers~\cite{GKR,GM,GKR2} in which the $\mu$-deformation based
approach was initiated and developed.

For our purposes we define the $\mu$-deformed trigonometric function
(see \cite{GKK,GKR2} and references therein) as
\begin{equation}
\sin_{\mu}{x}=\sum\limits_{n=0}^\infty (-1)^n\frac{x^{2n+1}}{[2n+1]_\mu!},\quad
\cos_{\mu}{x}=\sum\limits_{n=0}^\infty (-1)^n\frac{x^{2n}}{[2n]_\mu!},
\end{equation}
where $[n]_\mu!=[1]_\mu\,[2]_\mu\ldots[n]_\mu$. Clearly, at $\mu\to
0$ one recovers customary sine and cosine.

For our purposes, we introduce the $\mu$-deformed analogs of
spherical Bessel functions, namely
\begin{equation}\label{base}
j^{(\mu)}_0(x)=\frac{\sin_{\mu}{x}}{x},\qquad
y^{(\mu)}_0(x)=\frac{\cos_{\mu}{x}}{x} .
\end{equation}
At $\mu=0$ these reduce to the familiar Bessel functions.

The physical motivation for introducing these functions is
two-fold: the first one in \textbf{Eq.~\ref{base}} describes
the density profile of the dark matter halo and also leads to the
rotation curves within the $\mu$-deformed extension \cite{GKK} of
the Bose-condensate model, while both functions, taken jointly, are of
importance for constructing the representation space of the position
and momentum operators, see \textbf{Sections~\ref{Sec3.1}--\ref{Sec3.2}} below.

\begin{figure}[th!]
\begin{center}
\includegraphics[width=16.1cm]{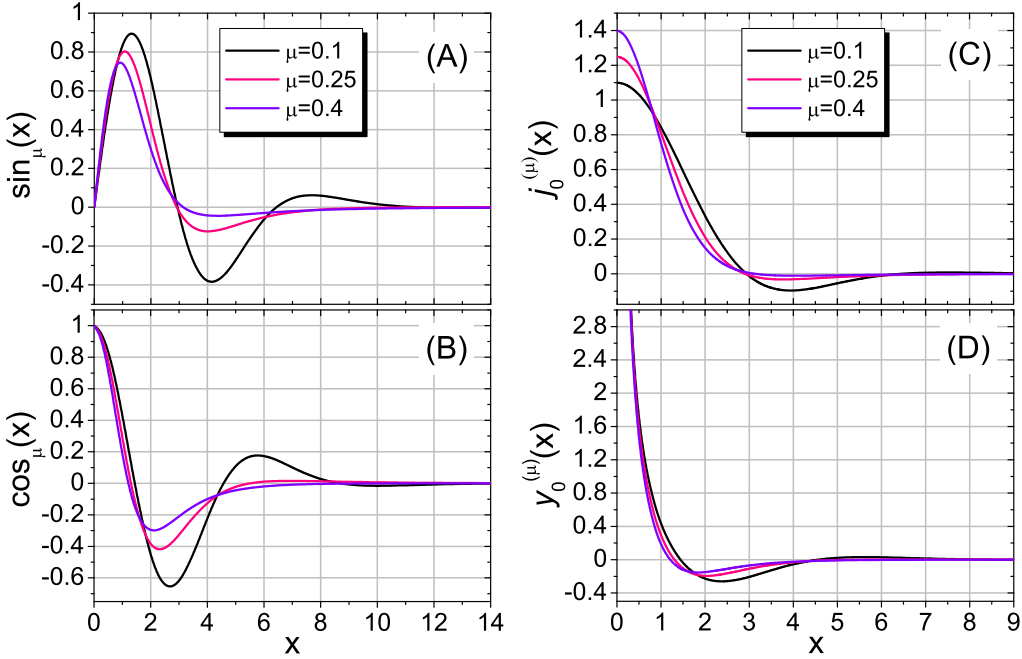}
\end{center}
\vspace*{-15mm}
\caption{Basic $\mu$-deformed trigonometric (\textbf{A} and \textbf{B})
and spherical (\textbf{C} and \textbf{D}) functions.}\label{fig1}
\end{figure}

Since the applied deformation concerns mainly the basic
trigonometric functions, let us study $\sin_{\mu}{x}$ and
$\cos_{\mu}{x}$ in detail. Contracting the corresponding series to
the Gaussian hypergeometric function, we can then represent them in
the analytic form
\begin{equation}\label{trig1}
\sin_{\mu}{x}=I(x,\mu)\sin{\varphi(x)},\qquad
\cos_{\mu}{x}=I(x,\mu)\cos{\varphi(x)}
\end{equation}
where
\begin{equation}\label{trig2}
I(x,\mu)\equiv(1+\mu^2x^2)^{-\frac{1+\mu}{2\mu}},\qquad
\varphi(x)\equiv\frac{1+\mu}{\mu}\arctan{(\mu x)}.
\end{equation}

Therefore, in the case of $\mu$-deformation, the main trigonometric
identity is written as follows:
\begin{equation}
\sin^2_{\mu}{x}+\cos^2_{\mu}{x}=I^2(x,\mu),\qquad I(x,\mu)\leq1.
\end{equation}
The behavior of the $\mu$-deformed trigonometric and spherical
functions is shown in \textbf{Figure~\ref{fig1}}.

In principle, it is possible to express the deformed trigonometric functions
in terms of the $\mu$-deformed exponential function. The $\mu$-analogs of
exponential and logarithmic functions are
\begin{equation}
\rme_\mu(x)=(1-\mu x)^{-\frac{1+\mu}{\mu}},\qquad
\ln_\mu(x)=\frac{1}{\mu}\left(1-x^{-\frac{\mu}{1+\mu}}\right),
\end{equation}
which give us the known functions at $\mu\to0^{-}$ due to the asymptotic formulas:
\begin{equation}
\rme(x)=\lim\limits_{n\to\infty}\left(1+\frac{x}{n}\right)^{n},\qquad
\ln(x)=\lim\limits_{n\to\infty}n\left(x^{1/n}-1\right).
\end{equation}

Note that the $\mu$-deformed functions exhibit a non-trivial property at $\mu>0$:
\begin{equation}
(\rme_\mu(x))^n=\rme_\mu\left(\frac{1-(1-\mu x)^n}{\mu}\right),\qquad
\ln_\mu(x^n)=\frac{1-(1-\mu \ln_\mu(x))^n}{\mu}.
\end{equation}

Focusing on the problems with spherical symmetry, we need to define
an inner product $\langle f|g\rangle$ in terms of which the real
functions $u_1(x)=j^{(\mu)}_0(x)$ and $u_2(x)=y^{(\mu)}_0(x)$ become
orthonormal on finite interval $x\in[0;R(\mu)]$:
\begin{equation}\label{base1}
\langle f|g\rangle=\int_0^{R(\mu)} f^*(x)g(x)\,w_\mu(x)\,\rmd x,\qquad
\langle u_i|u_j\rangle=\delta_{i,j},
\end{equation}
where the asterisk means complex conjugation; the Latin indexes $i$, $j$ run from 1 to 2.

For the orthogonality of $\sin{\varphi}$ and $\cos{\varphi}$ on the
interval $\varphi\in[0;\pi]$, we constitute {\it ad hoc}
\begin{equation}\label{mes1}
w_\mu(x)\,\rmd x=\frac{2}{\pi}\,x^2 I^{-2}(x,\mu)\,\rmd\varphi(x),\qquad \pi=\varphi(R(\mu)),
\end{equation}
and obtain
\begin{equation}\label{mes2}
w_\mu(x)=\frac{2(1+\mu)}{\pi}x^2(1+\mu^2x^2)^{\frac{1}{\mu}},\qquad
R(\mu)=\frac{1}{\mu}\tan{\frac{\mu\pi}{1+\mu}},
\end{equation}
where $R(\mu)$ coincides with the first zero of $\sin_{\mu}{x}$.

Expanding these as
\begin{eqnarray}
w_\mu(x)&=&\frac{2}{\pi}x^2\left[1+(1+x^2)\mu+\left(x^2+\frac{x^4}{2}\right)\mu^2+O(\mu^3)\right],\\
R(\mu)&=&\pi-\pi\mu+\left(\pi+\frac{\pi^3}{3}\right)\mu^2+O(\mu^3),
\end{eqnarray}
we see that the known quantities are restored at $\mu=0$.

\subsection{Deformed Differential Calculus}

We would like to define the $\mu$-deformed derivative $D^{(\mu)}_x$
with respect to the positive variable $x$, and its inverse. Let the
functions $f(x)$ and $\phi(x)$ admit expansion in the Taylor series
and satisfy the relation
\begin{equation}\label{diff}
D^{(\mu)}_xf(x)=\phi(x).
\end{equation}
The actions of $D^{(\mu)}_x$ and antiderivative
$\left(D^{(\mu)}_x\right)^{-1}$ are respectively given as
\begin{eqnarray}
\phi(x)&=&\frac{\rmd}{\rmd x}\left[f(x)
-x^{-\frac{1}{\mu}}\int_0^x f^\prime(s)\,s^{\frac{1}{\mu}}\,\rmd s\right],
\label{diff2}\\
f(x)&=&\mu x\,\phi(x)+\int_0^x \phi(s)\,\rmd s+f(0),
\end{eqnarray}
were the prime means ordinary differentiation.

We see that $\phi(x)=f^\prime(x)$ at $\mu\to0$ due to vanishing
$(s/x)^{1/\mu}$ for $s<x$. By definition, the derivative
$D^{(\mu)}_x$ lowers the exponent of the monomial $x^n$ by one,
namely $D^{(\mu)}_x x^n=[n]_\mu x^{n-1}$. However, the operator
$D^{(\mu)}_x$ violates the Leibniz rule: $D^{(\mu)}_x(f(x)\,
g(x))\not=g(x) D^{(\mu)}_xf(x)+f(x) D^{(\mu)}_xg(x)$.

\begin{table}[th!]
\renewcommand*{\arraystretch}{2.2}
\begin{center}
\begin{tabular}{|c|c|c|}
 \hline
   & $f(x)$ & $D^{(\mu)}_xf(x)$ \\ [0.5ex]
 \hline\hline
 1 & $x^n$ & $[n]_\mu x^{n-1}$ \\
 [1ex]\hline
 2 & $\rme_\mu(px)$ & $p\,\rme_\mu(px)$ \\
 [1ex]\hline
 3 & $\ln_\mu(x)$ & $(1+\mu-\mu^2)^{-1}x^{-2+\frac{1}{1+\mu}}$ \\
 [1ex]\hline
 4 & $\sin_\mu(x)$ & $\cos_\mu(x)$ \\
 [1ex]\hline
 5 & $\cos_\mu(x)$ & $-\sin_\mu(x)$ \\
 [1ex]\hline
 6 & $j^{(\mu)}_0(x)$ & $\frac{1+\mu}{1-\mu}\,y^{(\mu)}_0(x)-\frac{1-\mu^2x^2}{(1-\mu)x}\,j^{(\mu)}_0(x)$ \\
 [1ex]\hline
 7 & $y^{(\mu)}_0(x)$ & $-\frac{1+\mu}{1-\mu}\,j^{(\mu)}_0(x)-\frac{1-\mu^2x^2}{(1-\mu)x}\,y^{(\mu)}_0(x)$ \\
 [1ex]\hline
\end{tabular}
\end{center}
\vspace*{-5mm}
\caption{The $\mu$-deformed derivatives.}\label{tab1}
\end{table}

The $\mu$-deformed derivatives of some functions are collected in \textbf{Table~\ref{tab1}}.
To derive expressions 4--7, we have used the known auxiliary integrals:
\begin{equation}
\int \sin^{p-1}{x}\left\{
\begin{array}{c}
\sin{((p+1)x)}\\
\cos{((p+1)x)}
\end{array}
\right\}\rmd x=\frac{1}{p}\sin^p{x}\left\{
\begin{array}{c}
\sin{(px)}\\
\cos{(px)}
\end{array}\right\}.
\end{equation}

On the base of relations 4--5 (not 6--7) in \textbf{Table~\ref{tab1}}, we define
the Hermitian momentum operator ${\hat P}$ as
\begin{equation}\label{mom}
{\hat P}=-\frac{\rmi}{x} D^{(\mu)}_xx,
\end{equation}
so that $\langle u_i|{\hat P}|u_i\rangle=0$, and $\langle u_1|{\hat P}|
u_2\rangle= \langle u_2|{\hat P}^*|u_1\rangle=\rmi$ (see
\textbf{Eq.~\ref{base1}}), using the imaginary unit $\rmi$. This
operator will play an important role in the study of the deformed
Heisenberg algebra further on.

To demonstrate the action of ${\hat P}$ on some functions, note that
${\hat P}x^n=-\rmi[n+1]_\mu x^{n-1}$ for $n\geq0$, and then
\begin{equation}
{\hat P}\psi_p(x)=p\psi_p(x),\qquad \psi_p(x)=\frac{\rme_\mu(\rmi px)}{x}.
\end{equation}

In addition, we consider the radial part $\Delta^{(\mu)}_r$ of
$\mu$-deformed Beltrami--Laplace operator and its inverse (up to the
additive constant $C\sim f(0)$):
\begin{eqnarray}
\Delta^{(\mu)}_rf(r)&\equiv&\frac{1}{r^2} D^{(\mu)}_r\left(r^2 D^{(\mu)}_rf(r)\right),
\label{Lap}\\
\left(\Delta^{(\mu)}_r\right)^{-1}f(r)&=&\mu^2r^2f(r)+(1+\mu)\int_0^r f(s)\,s\,\rmd s
-\frac{1-\mu}{r}\int_0^r f(s)\,s^2\,\rmd s+C.
\end{eqnarray}

It is easy to verify for positive $n$ that
\begin{equation}
\Delta^{(\mu)}_r r^n=[n]_\mu \cdot [n+1]_\mu\, r^{n-2};\qquad
\left(\Delta^{(\mu)}_r\right)^{-1} r^n=\frac{r^{n+2}}{[n+2]_\mu \cdot [n+3]_\mu},
\quad C=0.
\end{equation}

We also verify that
\begin{equation}
\left(\Delta^{(\mu)}_r\right)^{-1} j^{(\mu)}_0(r)+j^{(\mu)}_0(r)=0, \qquad C=-j^{(\mu)}_0(0),
\end{equation}
by the use of the integrals
\begin{equation}
\int \cos^{p-1}{x}\left\{
\begin{array}{c}
\sin{((p+1)x)}\\
\cos{((p+1)x)}
\end{array}
\right\}\rmd x=\frac{1}{p}\cos^p{x}\left\{
\begin{array}{c}
-\cos{(px)}\\
\sin{(px)}
\end{array}\right\}.
\end{equation}

\section{Deformed Heisenberg Algebra and Uncertainty Principle}

\subsection{Deformed Equations and Heisenberg Algebra}\label{Sec3.1}

Here we are going to present the equations of some models using
deformed differential calculus. The main model for us, from which
the deformed Heisenberg algebra will follow, is described by the
deformation of the Lane--Emden equation (LEE) for finite density
function $\rho(r)$ in the two possible formulations
\begin{eqnarray}
&&\left(\Delta^{(\mu)}_r\rho(r)+k^2\right)\rho(r)=0,\label{DMeq1}\\
&&\left(D^{(\mu)}_r D^{(\mu)}_r+g_\mu(r)\frac{2}{r}
D^{(\mu)}_r+h_\mu(r)k^2\right)\rho(r)=0,\label{DMeq2}
\end{eqnarray}
where
\begin{equation}
g_\mu(r)=\frac{1}{1-2\mu}\left(1-\frac{1-\mu}{1+\mu}\mu^2k^2r^2\right),\qquad
h_\mu(r)=\frac{1+2\mu}{1-2\mu}-2\mu^2\frac{1-\mu^2k^2r^2}{(1+\mu)(1-2\mu)}.
\end{equation}
Note that the version \textbf{Eq.~\ref{DMeq1}} of $\mu$-deformed LEE
was already dealt with earlier in~\cite{GKK}, whereas the version in
\textbf{Eq.~\ref{DMeq2}} is completely new, unpublished one. As
seen, $g_\mu(r)\to 1$ and $h_\mu(r)\to 1$ at $\mu\to0$.

It is important that, due to the special form of $g_\mu(r)$ and
$h_\mu(r)$, these two $\mu$-deformed versions of LEE have the same
physical solution $j^{(\mu)}_0(kr)$ (along with $y^{(\mu)}_0(kr)$)
at $\mu<0.5$, which means that the two versions are
equivalent. To display this equivalence we have to explicitly
transform \textbf{Eq.~\ref{DMeq1}} into \textbf{Eq.~\ref{DMeq2}}.
Setting $kr\equiv x$ for simplicity, we assume the permutation rule
as $D^{(\mu)}_x\,x=\sigma(x)\,x\, D^{(\mu)}_x+\lambda(x)$, apply it
twice to the operator $D^{(\mu)}_r\, r^2\, D^{(\mu)}_r$ of the
$\mu$-Laplace operator in \textbf{Eq.~\ref{Lap}}, and find the functions
$\sigma(x)$ and $\lambda(x)$. Then, the equivalence of \textbf{Eq.~\ref{DMeq1}}
and \textbf{Eq.~\ref{DMeq2}} is seen, with the nontrivial commutation relation:
\begin{eqnarray}
&&\sigma(x)\,x\, D^{(\mu)}_x-D^{(\mu)}_x\,x=-\lambda(x),\label{DHA}\\
&&\sigma(x)=\!\frac{1}{\sqrt{h_\mu}}\!=\!        
\left[\frac{(1-2\mu)(1+\mu)}{1\!+\!\mu(3\!+\!2\mu^3x^2)}\right]^{1/2},\quad
\lambda(x)=\frac{2g_\mu}{(1\!+\!\sigma)h_\mu}\!=\!\frac{1+\mu-(1\!-\!\mu)\mu^2x^2}
{\mu[2+\mu(1+\mu^2x^2)]}(1\!-\!\sigma).\label{DHAf}
\end{eqnarray}
As result, we have come to {\it the new ($\mu$-deformed)
generalization of Heisenberg algebra}.

\vspace*{-10mm}
\begin{figure}[th!]
\begin{center}
\includegraphics[width=16.1cm]{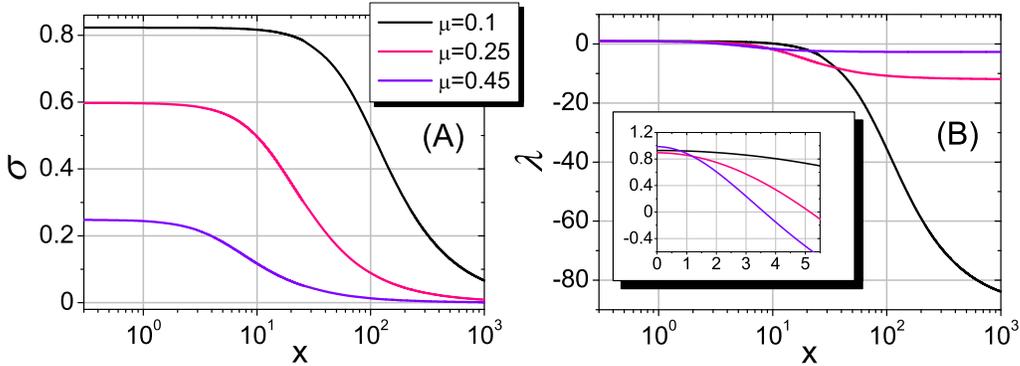}
\end{center}
\vspace*{-15mm}
\caption{Functions $\sigma(x)$ (\textbf{A}) and $\lambda(x)$ (\textbf{B})
in a wide range of variable $x$ and at fixed $\mu$.}\label{fig2}
\end{figure}

The functions $\lambda(x)$ and $\sigma(x)$ are real for $0<\mu<0.5$,
tend to 1 at $\mu\to0$, and are shown in \textbf{Figure~\ref{fig2}}.
We have $0\leq\sigma(x)\leq1$, while the maximum of $\lambda(x)$ is
determined by $\lambda(0)$ and is equal to
\begin{equation}
\lambda_{\rm max}(\mu)=\frac{1+\mu}{(2+\mu)\mu}\left(1-\sqrt{\frac{(1-2\mu)(1+\mu)}{1+3\mu}}\right).
\end{equation}
Although the function $\lambda(x)$ has a tail in negative values for
\begin{equation}
x>x_{\rm max}(\mu)=\frac{1}{\mu}\sqrt{\frac{1+\mu}{1-\mu}},
\end{equation}
as shown in \textbf{Figure~\ref{fig2}}, consideration of the problem
over finite interval of $x\in[0;R(\mu)]$ with $R(\mu)\leq x_{\rm
max}(\mu)$ for $\mu\in(0;0.5]$ guarantees a positive value of
$\lambda(x)$. Therefore, $R(\mu)$ varies between $R_{\rm
min}\simeq2.886$ and $R_{\rm max}=2\sqrt{3}\simeq3.464$, and it is
the minimum positive number that satisfies the condition
$\sin_\mu{R(\mu)}=0$ (see \textbf{Eq.~\ref{mes2}}).

It seems to be of interest to consider, elsewhere, the
quantum-mechanical problem of the propagation of a particle,
viewed as a spherical wave $\Psi(r)$ in a space curved due to
$\mu$-deformation. Without specifying the boundary condition, it can
be formulated as follows:
\begin{equation}\label{pcan}
\hat{P}^2\Psi(r)=k^2\Psi(r),
\end{equation}
where the momentum operator \textbf{Eq.~\ref{mom}} for $x=r$ is used.
Let us remark again that the operator $D^{(\mu)}_r$ in $\hat{P}$ is
a pseudohermitian one, see e.g.~\cite{Mos,BF,GK3,GK4}, but the ``sandwiching''
$\eta^{-1}D^{(\mu)}_r \eta$ with $\eta=r$ transforms it into
Hermitian form as in \textbf{Eq.~\ref{pcan}}.

In view of the definition \textbf{Eq.~\ref{mom}} of the momentum
operator, we formulate our $\mu$-deformed Heisenberg algebra:
\begin{equation}\label{DHA2}
\sigma(x)\,x\,{\hat P}-{\hat P}\,x=\rmi\lambda(x).
\end{equation}
In what follows, we will focus on the study of the uncertainty
principle (relation) which follows from the algebra
\textbf{Eq.~\ref{DHA2}}.

\subsection{Generalized Uncertainty Principle}\label{Sec3.2}

Denoting the standard deviations as
\begin{equation}
\Delta x=\sqrt{\langle x^2\rangle-\langle x\rangle^2},\qquad
\Delta P=\sqrt{\langle {\hat P}^2\rangle-\langle {\hat P}\rangle^2},
\end{equation}
we proceed to the analysis of the generalized uncertainty relation (GUP)
\begin{equation}\label{GUP}
\Delta x\,\Delta P\geq\frac{1}{2}|\langle[x,{\hat P}]\rangle|,
\end{equation}
where the commutator is taken from \textbf{Eq.~\ref{DHA2}}.

To gain insight into the general properties of
\textbf{Eq.~\ref{GUP}} for the $\mu$-deformed Heisenberg algebra
\textbf{Eq.~\ref{DHA2}}, let us combine \textbf{Eq.~\ref{DHA2}} with
its Hermitian conjugate to obtain
\begin{equation}
[(1+\sigma(x))x,{\hat P}]=2\rmi\lambda(x).
\end{equation}

Applying \textbf{Eq.~\ref{GUP}} to this commutation relation, we get
\begin{equation}
\Delta[(1+\sigma)x]\,\Delta P\geq\langle\lambda(x)\rangle.
\end{equation}
Taking into account that $1\geq\sigma(x)>0$ for $\mu<0.5$ in the
left hand side, we come to the GUP
\begin{equation}\label{gup2}
\Delta x\,\Delta P\geq\frac{1}{2}\langle\lambda(x)\rangle.
\end{equation}

To evaluate the averages, we specify the states
similarly to quantum ones. So, let us consider a normalized mixed
state $|\xi\rangle$ for $\xi\in[0;2\pi)$ in a Hilbert space basis
\textbf{Eq.~\ref{base}} endowed with the inner product from
\textbf{Eq.~\ref{base1}}:
\begin{equation}\label{xi1}
|\xi\rangle=\cos{\xi}|u_1\rangle+\sin{\xi}|u_2\rangle.
\end{equation}
Here $u_1(x)=j^{(\mu)}_0(x)$ and $u_2(x)=y^{(\mu)}_0(x)$ as before.

In fact, the mixed state $|\xi\rangle$ represents a general solution
to the $\mu$-deformed LEE, given by \textbf{Eq.~\ref{DMeq1}} and
\textbf{Eq.~\ref{DMeq2}}. Since the $\mu$-deformed LEE is formulated
for the local density of matter and, therefore, basically differs from
the complex-valued Schr\"odinger equation, it is natural to describe its
solution from \textbf{Eq.~\ref{xi1}} in terms of real-valued functions.
Although the state $|0\rangle$ for $\xi=0$, such that
$\langle x|0\rangle=j^{(\mu)}_0(x)$, serves to describe the finite DM
distribution in \cite{GKK}, the case $\xi\not=0$ admits the contribution
of the cuspidal distribution $y^{(\mu)}_0(x)$ at $x\to0$.

Thus, we define the mean:
\begin{equation}\label{qmean}
\langle(\ldots)\rangle=\langle\xi|(\dots)|\xi\rangle
\end{equation}
for fixed $\xi\in[0;2\pi]$ and $0<\mu<0.5$.

In contrast to quantum mechanics, \textbf{Eq.~\ref{qmean}} suggests
to evaluate a mean of some operator $(\ldots)$ in the basis generated
by the $\mu$-deformed LEE. There is no mathematical incorrectness in
choosing basis functions coinciding with physical distributions. Only
in turning to an interpretation, does one face the averaging (of powers)
of the distribution function itself (this also happens in multifractal
analysis).

The necessary matrix elements are given by
\begin{eqnarray}
&&\langle\xi|f(x)|\xi\rangle=A_f(\mu)-B_f(\mu)\cos{(2\xi)}+C_f(\mu)\sin{(2\xi)},\\
&&\langle\xi|f(x)|\xi+\pi/2\rangle=B_f(\mu)\sin{(2\xi)}+C_f(\mu)\cos{(2\xi)},\\
&&\left\{
\begin{array}{c}
A_f\\
B_f\\
C_f
\end{array}\right\}=\frac{1}{\pi}\int_0^\pi
\left\{
\begin{array}{c}
1\\
\cos{(2\varphi)}\\
\sin{(2\varphi)}
\end{array}\right\}\,f(X(\varphi))\,\rmd\varphi,\quad
X(\varphi)=\frac{1}{\mu}\tan{\frac{\mu\varphi}{1+\mu}}.
\end{eqnarray}

It is immediately seen that
\begin{equation}
\langle{\hat P}\rangle=\langle\xi|{\hat P}|\xi\rangle=-\rmi\langle\xi|\xi+\pi/2\rangle=0,\quad
\langle{\hat P}^2\rangle=\langle\xi|{\hat P}^2|\xi\rangle=-\langle\xi|\xi+\pi\rangle=1,
\end{equation}
Therefore, the standard deviation $\Delta P=1$ is fixed for the set of states $\{|\xi\rangle\}$.

On the other hand, let us introduce the functions
\begin{equation}\label{Ll}
\Delta x(\xi,\mu)\equiv\sqrt{\langle\xi|x^2|\xi\rangle-(\langle\xi|x|\xi\rangle)^2},\quad
\Lambda(\xi,\mu)\equiv\langle\xi|\lambda(x)|\xi\rangle,
\end{equation}
which represent the averages $\Delta x$ and $\langle\lambda(x)\rangle$, respectively.

\vspace*{-13mm}
\begin{figure}[th!]
\begin{center}
\includegraphics[width=16.1cm]{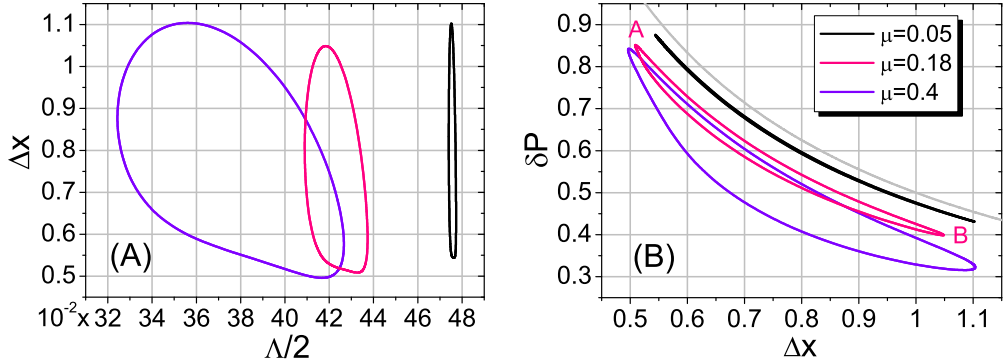}
\end{center}
\vspace*{-15mm}
\caption{\textbf{(A)}: Position deviation $\Delta x(\xi,\mu)$ versus mean
$\langle\lambda(x)\rangle=\Lambda(\xi,\mu)$, which are computed in the state
$|\xi\rangle$ at fixed $\mu$. \textbf{(B)}: Limiting momentum deviation
$\delta P(\xi,\mu)=\Lambda/(2\Delta x)$ versus $\Delta x(\xi,\mu)$. Turning
points of the pink banana-like curve are $A(0.51;0.85)$, $B(1.05;0.4)$.
Grey line corresponds to $\delta P=1/(2\Delta x)$.}\label{fig3}
\end{figure}

Thus, in the basis of the $\mu$-deformed spherical waves, one has
$\Delta P=1$, and it is required that
\begin{equation}\label{gup3}
\Delta x(\xi,\mu)\geq\frac{1}{2}\Lambda(\xi,\mu).
\end{equation}
This relation can be analyzed with the help of \textbf{Figure~\ref{fig3}A}.

Let us introduce the auxiliary momentum variance, accounting for \textbf{Eq.~\ref{gup3}}:
\begin{equation}\label{nP}
\delta P(\xi,\mu)=\frac{\Lambda(\xi,\mu)}{2\Delta x(\xi,\mu)}\leq1.
\end{equation}
We see that $\delta P(\xi,\mu)\leq\Delta P=1$ and
$\delta P(\xi,\mu)\,\Delta x(\xi,\mu)=\Lambda(\xi,\mu)/2$ by definition.
The behavior of $\delta P(\xi,\mu)$ is shown in \textbf{Figure~\ref{fig3}B}.

\vspace*{-5mm}
\begin{figure}[th!]
\begin{center}
\includegraphics[width=8.1cm]{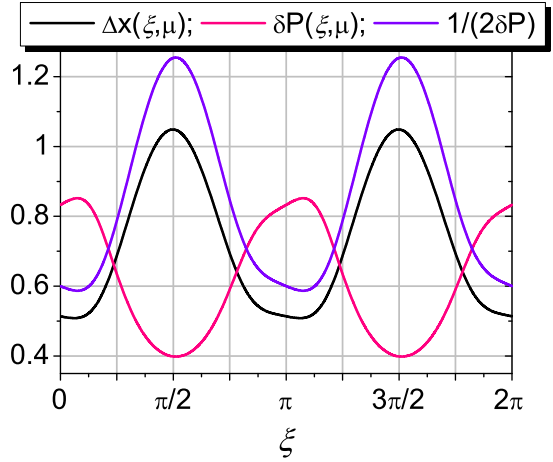}
\end{center}
\vspace*{-13mm}
\caption{Dependence of the lower limit deviations
$\Delta x(\xi,\mu)$ and $\delta P(\xi,\mu)$ on the state labeled by
$\xi$ at $\mu=0.18$. The running deviations $\Delta x$ and $\Delta P$
satisfy the GUP $\Delta x\,\Delta P\geq\Lambda/2$ and vary in the ranges
$\Delta x(\xi,\mu)\leq\Delta x\leq R(\mu=0.18)\simeq2.89$ and
$\delta P(\xi,\mu)\leq\Delta P\leq 1$.}\label{fig4}
\end{figure}

The admissible domain of variety of the running values of $\Delta x$ and $\Delta P$
is shown in \textbf{Figure~\ref{fig4}}. We see that the black and pink curves are
in antiphase regime, as it should be. For comparison, the violet curve describes
the change in the deviation $\Delta x$ according to the hyperbolic law in
accordance with the standard Heisenberg algebra.

\subsubsection{Alternative Approach}

To confirm the validity of \textbf{Eq.~\ref{gup2}} for the algebra
\textbf{Eq.~\ref{DHA2}} non-linear in $x$, it is worth to develop an
alternative calculation scheme applicable to various ways of writing
the commutator for $x$ and ${\hat P}$. For instance, there is a
possibility to rewrite relation \textbf{Eq.~\ref{DHA2}} in
equivalent form as
\begin{equation}\label{DHA3}
[x,{\hat P}]=\rmi\frac{2\lambda(x)}{1+\sigma(x)}+\frac{1-\sigma(x)}{1+\sigma(x)}\{x,{\hat P}\},
\end{equation}
where $\{x,{\hat P}\}\equiv x{\hat P}+{\hat P}x$ is anticommutator.

To analyze the GUP given by \textbf{Eq.~\ref{GUP}} for this commutation relation,
we assume that the brackets $\langle(\ldots)\rangle$ mean the quantum average over
the state defined by the {\it real} wave function in the coordinate representation.
Then, the action of the operator $\rmi{\hat P}$ on such a state results in a real-valued
expression, what immediately yields
\begin{equation}
|\langle[x,{\hat P}]\rangle|=2\left\langle\frac{\lambda(x)}{1+\sigma(x)}\right\rangle
-\left\langle\frac{1-\sigma(x)}{1+\sigma(x)}\{x,\rmi{\hat P}\}\right\rangle,
\end{equation}
when the positive first term on the right hand side dominates the second one.

In view of the inequality $|\langle{\hat A}{\hat
B}\rangle|\leq|\langle{\hat A}\rangle|\, |\langle{\hat B}\rangle|$,
we split the last term as
\begin{equation}
\left|\left\langle\frac{1-\sigma(x)}{1+\sigma(x)}\{x,\rmi{\hat P}\}\right\rangle\right|\leq
\left\langle\frac{1-\sigma(x)}{1+\sigma(x)}\right\rangle\,|\langle\{x,\rmi{\hat P}\}\rangle|,\quad
0\leq\sigma(x)\leq1.
\end{equation}

At this stage, we obtain
\begin{equation}
|\langle[x,{\hat P}]\rangle|\geq 2\left\langle\frac{\lambda(x)}{1+\sigma(x)}\right\rangle
-\left\langle\frac{1-\sigma(x)}{1+\sigma(x)}\right\rangle\,|\langle\{x,\rmi{\hat P}\}\rangle|.
\end{equation}

To evaluate $|\langle\{x,\rmi{\hat P}\}\rangle|$, we introduce the
operators $\delta x=x-\langle x\rangle$ and $\delta{\hat P}={\hat
P}-\langle{\hat P}\rangle$, where the hat over ${\hat P}$
distinguishes the operator $\delta{\hat P}$ from the function
$\delta P$ in \textbf{Eq.~\ref{nP}}. Then one gets
\begin{equation}
\langle\{x,\rmi{\hat P}\}\rangle=2\rmi\langle x\rangle \langle{\hat P}\rangle
+\langle\{\delta x,\rmi\delta{\hat P}\}\rangle.
\end{equation}

Due to the Cauchy--Schwartz inequality $|\langle{\hat A}{\hat B}\rangle|^2\leq|\langle{\hat A}^2\rangle|\,|\langle{\hat B}^2\rangle|$,
the following estimate holds:
\begin{equation}
|\langle\{\delta x,\rmi\delta{\hat P}\}\rangle|\leq2\Delta x\,\Delta P.
\end{equation}

Since the dark matter flux is assumed to be absent in the halo, one
can put $\langle{\hat P}\rangle=0$, which is confirmed by our direct
calculations. Combining, we have the estimate
\begin{equation}
|\langle[x,{\hat P}]\rangle|\geq 2\left\langle\frac{\lambda(x)}{1+\sigma(x)}\right\rangle
-2\left\langle\frac{1-\sigma(x)}{1+\sigma(x)}\right\rangle\,\Delta x\,\Delta P.
\end{equation}

Substituting it into \textbf{Eq.~\ref{GUP}} and accounting for $\langle1\rangle=1$,
we arrive at
\begin{eqnarray}
&&\Delta x\,\Delta P\geq \frac{1}{2}\langle\lambda(x)\rangle_W,\label{gup4}\\
&&\langle(\dots)\rangle_W\equiv\frac{\langle W(x)(\ldots)\rangle}{\langle W(x)\rangle},\quad
W(x)=\frac{1}{1+\sigma(x)},\label{Wav}
\end{eqnarray}
where the new mean $\langle(\dots)\rangle_W$ with additional convex
weighting function $W(x)$ arises.

For a given function $W(x)$ we get
\begin{equation}
\langle\lambda(x)\rangle_W=\langle\lambda(x)\rangle+\frac{\langle\delta W(x) \delta\lambda(x)\rangle}{\langle W(x)\rangle},
\end{equation}
where $\langle\delta W(x) \delta\lambda(x)\rangle$ is a covariance between
the convex function $W(x)$ and concave $\lambda(x)$, and it is determined
by deviations like $\delta f(x)=f(x)-\langle f(x)\rangle$.

Since the function $\sigma(x)$ (and $W(x)$) changes only slightly
over the interval $x\in[0;R(\mu)]$ in \textbf{Figure~\ref{fig2}A},
it can be approximated by a constant close to $\sigma(0)$ when
calculating integrals. This provides $\delta W(x)\to0$ and
numerically leads to expressions:
\begin{equation}\label{aav}
\left\langle\frac{\lambda(x)}{1+\sigma(x)}\right\rangle\simeq
\frac{\langle\lambda(x)\rangle}{1+\langle\sigma(x)\rangle},\qquad
\left\langle\frac{1}{1+\sigma(x)}\right\rangle\simeq
\frac{1}{1+\langle\sigma(x)\rangle},
\end{equation}
when we use $\langle(\ldots)\rangle=\langle\xi|(\ldots)|\xi\rangle$ in the range $0<\mu<0.5$.

This circumstance leads again to \textbf{Eq.~\ref{gup2}} for the
states $|\xi\rangle$, that is just \textbf{Eq.~\ref{gup3}}.

Note that the appearance of the mean \textbf{Eq.~\ref{Wav}} is
associated with the initial \textbf{Eq.~\ref{DHA3}} for the
commutation relation. In other cases, we may only encounter means of
type \textbf{Eq.~\ref{aav}}, where it would be justified to use the
estimate $1\geq\langle\sigma(x)\rangle$.

\subsection{Application to Dark Matter}\label{Sec3.3}

Let us remind the connection between the operators in terms of the
dimensionless variable $x=kr$ and the operators of the physical
radial coordinate $r$ and the momentum ${\hat P}_r$:
\begin{equation}
r=\frac{x}{k},\qquad {\hat P}_r=\hbar k{\hat P},
\end{equation}
where $k$ is the parameter of \textbf{Eq.~\ref{DMeq1}, \ref{DMeq2}}
and has the dimension of inverse length; the operator ${\hat P}$ is
given by \textbf{Eq.~\ref{mom}}.

\begin{table}[h!]
\renewcommand*{\arraystretch}{2.2}
\begin{center}
{\footnotesize
\begin{tabular}{|c|c|c|c|c|c|c|}
 \hline
  Galaxy & $\mu$ & $k$, ${\rm kpc}^{-1}$ & $(\Delta r)_{\rm max}$, ${\rm kpc}$ & $(\Delta r)_{\rm min}$, ${\rm kpc}$ &
    $(\Delta P_r)_{\rm max}$, $10^{-27}~{\rm eV}/c$ & $(\Delta P_r)_{\rm min}$, $10^{-27}~{\rm eV}/c$ \\ [0.5ex]
 \hline\hline
 M81dwB & $0.18$ & $2.64$ & $0.398$  & $0.193$ & $14.38$ & $6.75$ \\
 [1ex]\hline
 DDO 53 & $0.18$ & $0.97$ & $1.082$  & $0.526$ & $5.28$ & $2.48$ \\
 [1ex]\hline
 IC 2574 & $0.179$ & $0.17$ & $6.18$  & $3.0$ & $0.926$ & $0.435$ \\
 [1ex]\hline
 NGC 2366 & $0.178$ & $0.37$ & $2.84$  & $1.38$ & $2.02$  & $0.946$ \\
 [1ex]\hline
 HO I & $0.151$ & $1.27$ & $0.830$  & $0.402$ & $6.98$ & $3.33$ \\
 [1ex]\hline
\end{tabular}
}
\end{center}
\caption{The parameters for the dark matter halos of dwarf galaxies.}\label{tab2}
\end{table}

The most successful results of paper~\cite{GKK} for describing the dark matter halo
of dwarf galaxies based on the $\mu$-deformed Lane--Emden equation were obtained
in the following range of parameters:
\begin{equation}
\mu=0.151\ldots0.18,\qquad k=0.17\ldots2.64~{\rm kpc}^{-1}.
\end{equation}

Using the turning points $A((\Delta x)_{\rm min}; (\delta P)_{\rm max})$ and
$B((\Delta x)_{\rm max}; (\delta P)_{\rm min})$ for fixed $\mu$ as in
\textbf{Figure~\ref{fig3}B}, we relate extreme physical values $\Delta r$ and
$\Delta P_r$ with dimensionless ones $\Delta x$ and $\delta P$ as
\begin{eqnarray}
&&\Delta r=\Delta x\,\left[\frac{k}{1~{\rm kpc}^{-1}}\right]^{-1}\,{\rm kpc},
\label{eq59}\\
&&\Delta P_r=\delta P\,\left[\frac{k}{1~{\rm kpc}^{-1}}\right]\times6.394\times10^{-27}~\frac{{\rm eV}}{c}.
\label{eq60}
\end{eqnarray}
The calculation results are collected in \textbf{Table~\ref{tab2}}.
Therein, we present the obtained data for five dwarf galaxies (from
the eight ones given in Table~1 of \cite{GKK}), because just for
these galaxies the $\mu$-deformation based description of the
rotation curves is most successful with respect to
earlier approaches, as it provides the best agreement with
observational data (certainly better then if one uses the profile
from the usual Bose-condensate model of DM being the solution of
non-deformed Lane--Emden equation as in~\cite{Harko}, or uses the
famous Navarro--Frenk--White profile~\cite{NFW}). Of course, the
remaining three galaxies can also be considered, but the choice of
five ones is both sufficient, trustful, and best suited for our
treatment and conclusions.

Note that both the Lane--Emden equation and its $\mu$-deformed
extensions determine the distribution function $\rho(r)$ of the
matter, not the wave function of single particle. Therefore, the
mean \textbf{Eq.~\ref{qmean}} is a quadratic form in the
distribution, related here with $|0\rangle$ which differs by a
multiplicative constant defining $\rho(0)$ \cite{GKK}. Generally
speaking, the state $|0\rangle$ may not determine the turning points
of a banana-like curve in the $(\Delta x,\delta P)$ plane in
\textbf{Figure~\ref{fig3}B}, along with the extreme values of the
deviations $\Delta r$ and $\Delta P_r$. Nevertheless, the
mathematically correct mean \textbf{Eq.~\ref{qmean}} can be used to
obtain new additional information about dark matter, even by means
of considering the moments $\langle\rho^n\rangle$ of the
distribution similarly to multifractal analysis. Besides, the
extreme deviations set the limits for the fluctuations of physical
quantities at $\xi=0$.

Without a deep study of the structure of averages here, let us
analyze the physical consequences of the data in
\textbf{Table~\ref{tab2}}. We see that in the non-relativistic theory
the momentum deviation $\Delta P_r=m\,\Delta\upsilon_r$, where $m$
is the particle mass, $\Delta\upsilon_r$ is the deviation of
particle radial velocity $\upsilon_r$. Since the original
work~\cite{GKK} was using bosons with $m\sim10^{-22}~{\rm eV}/c^2$,
we obtain from \textbf{Table~\ref{tab2}} that
$\Delta\upsilon_r\sim10^{-5}c$ in units of the speed of light $c$.
Moreover, deviation of the kinetic energy $\Delta
E_K=m(\Delta\upsilon_r)^2/2$ can be used to determine the effective
temperature of dark matter, namely $T_{\rm eff}=(\Delta
P_r)^2/(2m)\sim10^{-32}~{\rm eV}$. This value is much smaller than
the critical temperature of the Bose--Einstein condensation, as it
should be in such a paradigm.

Due to the GUP given by \textbf{Eq.~\ref{gup2}}, we relate the temperature $T_{\rm eff}$
of the spherical layer in the vicinity of $\langle r\rangle$ to its width $2 \Delta r$ :
\begin{equation}\label{rTL}
(\Delta r)^2\,T_{\rm eff}\geq\frac{\hbar^2}{8m}\,\langle\lambda\rangle^2.
\end{equation}
This formula holds for a macroscopic system of finite volume
when $\Delta r$ does not exceed the radius of the system, and it
shows that a smaller domain may have a higher temperature,
and vice verse.

It is worth to note that the mean $\langle\lambda\rangle$ in
\textbf{Eq.~\ref{rTL}} takes values in the limited interval
$\langle\lambda\rangle=\Lambda(\xi,\mu)\in[\Lambda_{\rm
min},\Lambda_{\rm max}]$, where the positive $\Lambda_{\rm min}$ and
$\Lambda_{\rm max}$ depend on $\mu$ (see \textbf{Eq.~\ref{Ll}} and
\textbf{Figure~\ref{fig3}A}). For $\mu=0.18$, we have $\Lambda_{\rm
min}\simeq0.818$ and $\Lambda_{\rm max}\simeq0.875$.

\section{Concluding Remarks}

In this work we have studied unusual consequences of the new
($\mu$-deformed) generalization of the Heisenberg algebra
\textbf{Eq.~\ref{DHA}, \ref{DHA2}}  which is
special as it was derived within the extension of Bose-condensate
dark matter model based on $\mu$-deformation. From the generalized
algebra we obtained nontrivial GUR that generates minimal and
maximal uncertainties of both positions (minimal/maximal lengths)
and momenta. The obtained GUR is strictly dependent on the states
(labeled by $\xi$) of the system, and such dependence was exploited
to full extent.

In \textbf{Table~\ref{tab2}}, the galaxies M81dwB and IC~2574 look
as the two ``extreme'' cases. Namely, for the latter we have the
largest $(\Delta r)_{\rm max}$ and $(\Delta r)_{\rm min}$, while for
the former these quantities show smallest values. Clearly, the
situation concerning $(\Delta P_r)_{\rm max}$ and $(\Delta P_r)_{\rm
min}$ is quite opposite. Noteworthy, the value of $\mu$ (strength of
deformation) for M81dwB and IC~2574 is almost the same. The
relations \textbf{Eq.~\ref{eq59}} and \textbf{Eq.~\ref{eq60}} show
the defining role of the quantity $k$ which involves scattering
length $a$ and particle mass $m$ as $k\propto m^{3/2} a^{-1/2}$ \cite{Harko}.

For the considered galaxies (each labeled by its specific value of $\mu$) we conclude:
since the particle mass is same (namely $10^{-22}~{\rm eV}/c^2$), we have differing scattering
lengths in halos of different galaxies (vice versa, would we assume same scattering length
for all the five galaxies we would have somewhat differing masses of DM particle in different galaxies,
though this second option seems to be less realistic).
As already shown in \cite{Harko}, where the BEC DM model is also based
on the LEE, there is no universality of model parameters when
describing all admissible objects. In fact, this issue remains in
our model, which improves the previously fitted rotation curves by
including an additional parameter $\mu$. Physically, we can only
control the applicability conditions of our model: consider
DM-dominated dwarf galaxies leaving aside their rigid rotation,
which contributes to the distribution function~\cite{Harko18,Naz20}.
Therefore, giving clear physical meaning to differing scattering
lengths in halos of different galaxies remains an interesting task
for future study.

Note that the parameter $k$ is related to the observed
radius $r_{\rm gal}$ of the galactic halo by $kr_{\rm gal}=R(\mu)$,
where the right-hand side is determined by the parameter $R(\mu)$
from \textbf{Eq.~\ref{mes2}}, replacing $\pi=R(0)$ in the
non-deformed case. We can easily find a small difference (of several
percent) between the values of $r_{\rm gal}$ in the deformed and
non-deformed cases, by comparing these with the galaxy radii from
\cite{Harko}. However, the simulation of rotation curves is  
more successful in the deformed case, as shown in~\cite{GKK}.

It is worth to remark that the values of $(\Delta r)_{\rm max}$ and
$(\Delta r)_{\rm min}$ for the two galaxies M81dwB and IC~2574, and
the others in \textbf{Table~\ref{tab2}}, reside well within the
observed sizes of DM halos as it should. Accordingly, the values of
$(\Delta P_r)_{\rm max}$ and $(\Delta P_r)_{\rm min}$ for these same
galaxies lie in the ranges completely consistent with DM being in
the ($\mu$-Bose) condensate state. Clearly this is in agreement with
the above reasonings concerning the effective temperature.

It is of interest to analyze possible special meaning of our results
on the existence of finite     
$(\Delta r)_{\rm max}$ and $(\Delta r)_{\rm min}$    
in the context of treatment in~\cite{LL,Lee} of minimum length scale
of galaxies (note that for the candidate length scales one can take
into consideration such concepts as coherence length, Compton
wavelength, quantum Jeans length scale, gravitational Bohr radius,
and de Broglie wavelength, see~\cite{LL} and references therein).
Time dependence of some of these quantities, e.g. characteristic
length scale $\tilde{\xi}$ (minimum size of DM dominated galaxies)
is studied in~\cite{Lee}. Let us quote one of the interesting
predictions of this work: with the mass of DM particles chosen as
$m=5\times 10^{-22}~{\rm eV}/c^2$, it follows that
$\tilde{\xi}(z=0)=311.5$~pc while $\tilde{\xi}(z=5)=81.2$~pc, i.e.
early dwarf galaxies were significantly more compact. In view of the
extremely tiny mass of the particle from dark sector, a question may
arise of possible (inter)relation of this entity with the cosmic
microwave background (CMB). The very first answer which comes to
one's mind could be that no relation is possible, because of the
absence of interaction between visible and dark sectors. However,
when considered in the framework of doubly special relativity, the
properties of the photon gas at these special conditions can appear,
see~\cite{CGN}, much more interesting and nontrivial. Noteworthy,
the treatment in~\cite{CGN}, on one hand, is potentially applicable
for studying some unclear features of CMB, and, on the other hand,
involves a kind of deformation which is very similar to the
$\mu$-deformation explored herein. We hope to address the details of
all these intriguing issues elsewhere.


\section*{Acknowledgment}
A.M.G. and A.V.N. acknowledge support from the National
Academy of Sciences of Ukraine by its Project No.~0122U000888.

\end{document}